\documentclass[aps,pra,twocolumn,superscriptaddress]{revtex4-2}
\usepackage[utf8]{inputenc}
\usepackage{soul}
\usepackage{textcomp}
\usepackage[english]{babel}
\usepackage{graphicx}
\usepackage{dcolumn}
\usepackage{bm}
\usepackage{amsmath}
\usepackage{verbatim}     
\usepackage[dvipsnames]{xcolor}       
\usepackage[normalem]{ulem}   
\usepackage{bbold}
\usepackage{mathrsfs}
\usepackage[mathscr]{eucal}
\usepackage{physics}
\usepackage{algorithmic}

\usepackage{hyperref}
\hypersetup{
   pdfpagemode=None,
   pdfstartpage=1,
   pdfmenubar=true,
   pdftoolbar=true,
   colorlinks = true,
   linkcolor=blue,
   citecolor=blue,
   urlcolor=blue,
   bookmarksopen=false
 } 

 \usepackage{booktabs}
 \usepackage{float}

\newcommand{\Cite}{\unskip~\cite}
\newcommand{\Eqref}{\unskip~\eqref}
\newcommand{\figref}{\unskip~\ref}
\newcommand{\qo}[1]{``#1''}

\begin{document}

\title{Single-acquisition tomography of photonic qubits with structured media}

\author{Francesco Di Colandrea}
\email{francesco.dicolandrea@unina.it}
\affiliation{Nexus for Quantum Technologies, University of Ottawa, K1N 5N6, Ottawa, ON, Canada}
\affiliation{Dipartimento di Fisica ``Ettore Pancini", Universit\`{a} degli Studi di Napoli Federico II, Complesso Universitario di Monte Sant'Angelo, Via Cintia, 80126 Napoli, Italy}

\author{Yingwen Zhang}
\affiliation{Nexus for Quantum Technologies, University of Ottawa, K1N 5N6, Ottawa, ON, Canada}
\affiliation{National Research Council of Canada, 100 Sussex Drive, K1A 0R6, Ottawa, ON, Canada}

\author{John Grace}
\affiliation{Nexus for Quantum Technologies, University of Ottawa, K1N 5N6, Ottawa, ON, Canada}
\affiliation{National Research Council of Canada, 100 Sussex Drive, K1A 0R6, Ottawa, ON, Canada}
\affiliation{McGill University, H3A 0G4, Montreal, QC, Canada}

\author{Dilip Paneru}
\affiliation{Nexus for Quantum Technologies, University of Ottawa, K1N 5N6, Ottawa, ON, Canada}
\affiliation{Dipartimento di Fisica ``Ettore Pancini", Universit\`{a} degli Studi di Napoli Federico II, Complesso Universitario di Monte Sant'Angelo, Via Cintia, 80126 Napoli, Italy}

\author{Alessio D'Errico}
\affiliation{Nexus for Quantum Technologies, University of Ottawa, K1N 5N6, Ottawa, ON, Canada}
\affiliation{National Research Council of Canada, 100 Sussex Drive, K1A 0R6, Ottawa, ON, Canada}

\author{Ebrahim Karimi}
\affiliation{Nexus for Quantum Technologies, University of Ottawa, K1N 5N6, Ottawa, ON, Canada}
\affiliation{National Research Council of Canada, 100 Sussex Drive, K1A 0R6, Ottawa, ON, Canada}
\affiliation{Institute for Quantum Studies, Chapman University, Orange, California 92866, USA}


\begin{abstract} 
Quantum state tomography is an essential tool for characterizing quantum systems and underpins nearly every experimental realization of quantum technologies. Conventional tomography relies on performing a sequence of projective measurements on many identical copies of a quantum state, requiring the measurement apparatus to be reconfigured between successive acquisitions. As the Hilbert-space dimension increases, the number of required measurements grows rapidly; in practice, additional overcomplete measurements are often performed to improve robustness to experimental imperfections. Here, we introduce a tomography platform based on structured anisotropic media that performs informationally complete measurements of photonic polarization qubits within a single acquisition. The approach employs three liquid-crystal metasurfaces with spatially varying optic-axis orientations that transform the input polarization into a far-field distribution of discrete transverse-momentum modes. Each diffraction pattern uniquely determines the polarization state, enabling its reconstruction without sequential changes to the measurement apparatus. Unlike previous implementations, our scheme is intrinsically photon-number independent: the same optical device operates identically for arbitrary photon numbers, while the desired photon-number sector can be selected afterwards through post-selection of the corresponding $n$-fold coincidence events. We experimentally demonstrate single-frame quantum state tomography of both single- and two-photon polarization states, providing a simple and scalable route toward efficient quantum-state characterization.
\end{abstract}

\maketitle


\section{Introduction}

Optical polarization has traditionally been used to encode two-level systems, as it is readily accessible in experimental settings. For this reason, the quantum state tomography of a qubit is often mapped onto the experimental characterization of the polarization state of a single photon \Cite{PhysRevA.64.052312,ALTEPETER2005105}. Accurate polarization characterization is key to probing the optical activity of structured materials \Cite{Tyo:06} or biological samples \Cite{Garcia:17,He2021}.

Conventional tomography is based on performing projective measurements of identically prepared qubits in different bases. A common choice of states is the set of mutually unbiased bases (MUBs), comprising eigenstates of the SU(2) generators. Within this decomposition, each pair of basis states yields the corresponding Stokes parameter, which provides the Cartesian representation of the qubit on the Bloch sphere. By adopting tensor products of Pauli matrices, this protocol easily generalizes to the multi-photon scenario \Cite{PhysRevA.64.052312}.

Despite being widely adopted in polarization tomography, MUBs provide a suboptimal choice, as they define an overcomplete set of six measurements. A convenient alternative is provided by the set of symmetric, informationally complete, positive operator-valued measure (SIC-POVM) states. These correspond to four projectors onto states associated with the vertices of a regular tetrahedron inscribed in the Bloch sphere, thereby defining a minimal tomographic set \Cite{10.1063/1.1737053,PhysRevX.5.041006}.

In conventional optical polarization tomography, each projection is implemented by adjusting the angles of a sequence of waveplates that, together with a polarizing filter, select the desired measurement basis. This requires modifying the measurement apparatus between successive realizations, which fragments the measurement process in time. In the context of foundational tests, such as violations of Bell inequalities, all joint correlation outcomes must be recorded simultaneously. Sequential projective measurements implemented by repeatedly adjusting waveplates cannot intrinsically provide this capability~\Cite{brunner2014bell}. Moreover, the number of measurements scales exponentially with the number of qubits, rapidly leading to prohibitive experimental complexity. It is therefore desirable to parallelize these operations to deliver a tomographic protocol that is, from an experimental perspective, single-acquisition. A straightforward solution is to fabricate compact metastructures in which different spatial sectors locally implement different polarization projections, enabling full or partial tomography \Cite{Hsu:14,Pors:15,Wen:15,Chen_2016,Wei:17,Arbabi2018FullStokes,10.1063/5.0102539,Soma:23,Fan2023,An2024}. 

Inspired by polarization imaging techniques \Cite{Solomon:81,Azzam:16}, a more refined strategy is to couple photons' polarization to their spatial structure via spin-orbit interactions of light \Cite{Bliokh2015,Cardano2015}, by routing each projection into distinct spatial channels \Cite{BalthasarMueller:16,EspinosaSoria2017,Yang2018,Dai:19,doi:10.1126/science.aat8196,Lung2024}. For example, such spatial multiplexing can be achieved by making the intensities associated with a set of transverse-momentum modes proportional to the success probabilities of a target set of tomographic projections. In this way, by implementing an optical Fourier transform, the intensities coupled to selected diffraction orders reveal in parallel the outcomes of the corresponding tomographic measurements. This concept underlies modern polarization-camera technologies \Cite{Yang2018,doi:10.1126/science.aax1839,Gottlieb:21,doi:10.1126/sciadv.adx9886,Bian2025}. The same strategy can be interpreted within the emerging framework of photonic reservoir computing, in which an unknown input is mapped into a high-dimensional feature space \Cite{Innocenti2023}. In this setting, the problem can also be naturally framed as a learning task for a neural network, which can be trained to infer the polarization state (or properties thereof) from the spatial structure of light emerging from a scattering medium \Cite{Pierangeli2023,pierangeli2} or from an ordered transformation like discrete-time quantum walks \Cite{PhysRevLett.132.160802,ziaScience}.

Along these lines, this work presents a spin-orbit gadget 
that maps $m$-photon polarization states onto a unique correlated distribution in the transverse-momentum degree of freedom, enabling complete state tomography within a single acquisition. The device comprises three liquid-crystal metasurfaces, designed such that different diffraction orders directly encode target polarization projections. Numerical optimization is needed only to ensure that the overall transformation is unitary. By design, the diffraction orders used for tomography are symmetric about the zero order, which enables multi-copy reconstructions to average over experimental imperfections. Unlike previous implementations, which require a number of interleaved metasurfaces that increases linearly with the number of photons \Cite{doi:10.1126/science.aat8196}, our platform always consists of three attached elements, independently of the number of input photons. In this sense, the platform is photon-number agnostic, and the number of photons to be characterized can also be selected directly in postprocessing. For single-photon states, the detection can be performed with a standard CCD/CMOS camera, while for multi-photon states, we assign different input qubits to distinct spatial positions along the direction perpendicular to the diffraction, and we extract the tomography from spatially resolved coincidence measurements using a time-stamping camera, i.e., TPX3CAM \Cite{Nomerotski_2023}. We further demonstrate the flexibility of the platform by implementing different tomographic sets of states, whereas previous methods based on linearly birefringent metagratings were limited to symmetry-constrained choices \Cite{doi:10.1126/science.aax1839}. 

In the following, we discuss the theoretical model of the measurement protocol, describe our design strategy, detail the optical implementation, and report experimental reconstructions of single- and two-photon polarization states.

\section{Theory}
\subsection{Coupling polarization to spatial modes}
The idea is to couple the two-dimensional polarization qubit space to a higher-dimensional Hilbert space spanned by the transverse modes of light. This interaction can be engineered via a unitary, space-dependent polarization transformation. Assuming that such a transformation is one-dimensional and periodic, it can be written as
\begin{equation}
U=\int_0^\Lambda \text{d}x \,\mathcal{U}(x)\ketbra{x},    
\label{eqn:unitary}
\end{equation}
where $x$ denotes the transverse coordinates (assuming that photons propagate along the $z$ axis), $\mathcal{U}(x)$ is an SU(2) operator corresponding to a local polarization transformation, and $\Lambda$ represents the spatial period: ${\mathcal{U}(x+\Lambda)=\mathcal{U}(x)}$. Crucially, due to its periodicity, $\mathcal{U}(x)$ can be decomposed into its discrete spatial frequencies at integer multiples of $2\pi/\Lambda$, by expressing the position states $\ket{x}$ as superpositions of momentum waves spanning the reciprocal space: ${\ket{x}=\sum_n e^{ixk_n}\ket{k_n}/\sqrt{\Lambda}}$, where ${k_n=2\pi n/\Lambda}$.  

\begin{figure*}[t!]
    \centering
    \includegraphics[width=\linewidth]{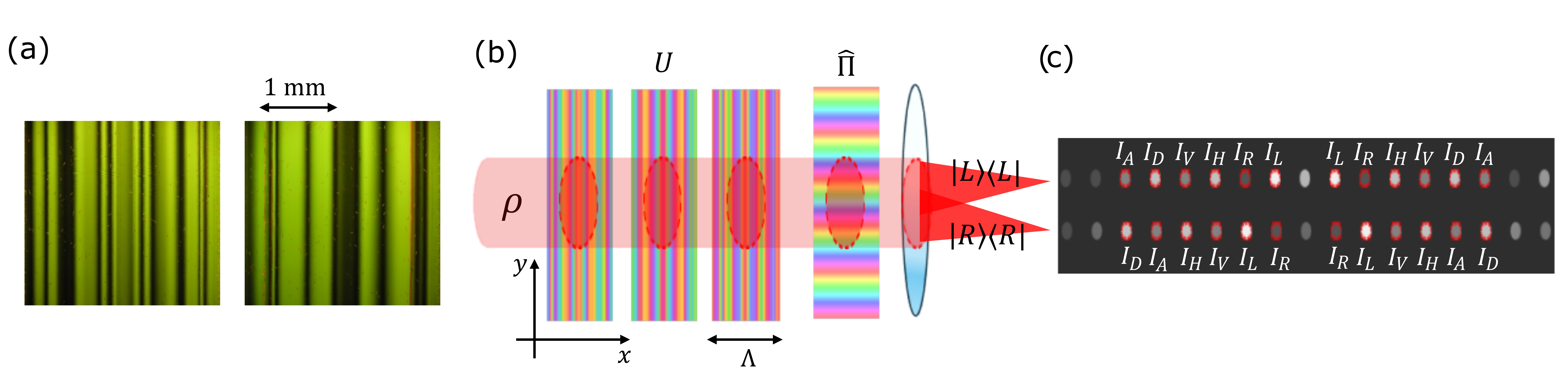}
    \vspace{-0.7cm}
    \caption{\textbf{Single-acquisition tomography with liquid-crystal metasurfaces.} (a) Patterns of two liquid-crystal metasurfaces for MUB and SIC-POVM tomography (left and right, respectively), viewed between crossed polarizers. Scale bars indicate the device dimensions. (b) An unknown polarization state, $\rho$, either pure or mixed, is coupled to transverse momentum through a space-dependent polarization transformation $U$, implemented with a fixed set of three liquid-crystal metasurfaces. (c) In the far field, upon a global polarization projection ${\hat{\Pi}}$, a target set of diffraction orders reveals the outcome of tomographic projections in parallel. For instance, in the MUB implementation, the measurements $\lbrace{I_H,I_V,I_A,I_D,I_L,I_R\rbrace}$ are spatially parallelized. If a polarization grating along $y$ is used, up to four equivalent tomographies can be extracted for the single-qubit case from a single experimental acquisition.}
    \label{fig:fig1}
\end{figure*}

Our platform is based on preparing the polarization qubit state, described by the density matrix $\rho$, in the central mode $ \ket{k_0}$, evolving the state under the periodic transformation, and eventually resolving the state in the transverse-momentum basis upon a global polarization projection. 
Experimentally, this corresponds to preparing $\rho$ in the fundamental Gaussian mode with beam waist ${w_0\geq\Lambda}$ \Cite{DErrico2020}, which we send through a set of three liquid-crystal metasurfaces, followed by a polarization projection. In the far field, i.e., in the focal plane of a lens, each discrete momentum mode $\ket{k_n}$ is focused into a different spot, occupying a distinct spatial position \Cite{goodman2005introduction}. The intensity of each spot is given by 
\begin{equation}\label{eqn:POVMprobabilities}
    p_n = \bra{\Pi, k_n}  U(\rho \otimes \ketbra{k_0}{k_0})U^\dagger\ket{\Pi, k_n}
    =\text{Tr}( E_n \rho ),
\end{equation}
where $\ket{\Pi}$ is the polarization state selected by the projection. Each spot corresponds to a positive operator-valued measure (POVM) element $E_n$,
\begin{equation}
    E_n = V_n^\dagger \ketbra{\Pi}{\Pi} V_n,
\end{equation}
where
\begin{equation}
V_n  =  \braket{k_n} {U|k_0}=\frac{1}{\Lambda}\int_0^\Lambda\text{d}x\,\mathcal{U}(x)e^{ixnk_0}.
\label{eqn:kraus}
\end{equation}
It follows that the set ${\{E_n\}\cup\{E_n^\perp\}}$ (where ${E_n^\perp}$ is defined analogously using the orthogonal state ${\ket{\Pi^\perp}}$) defines a POVM on the qubit space. 
POVMs form the foundation for generalized quantum measurements, playing a pivotal role in tomography. A POVM is said to be informationally complete if the measurement outcomes $\{p_n\}$, as defined in Eq.~\Eqref{eqn:POVMprobabilities}, uniquely determine the state $\rho$. In this paper, we describe a deterministic procedure for mapping a subset ${\lbrace{E_n: 1 \leq |n| \leq n_{\text{max}}\rbrace}}$ of our experimental POVM onto a target POVM:  
\begin{equation}\label{eqn:constraint}
    E_n = \lambda^2 E_n^{\text{(target)}}, 
\end{equation}
where $\lambda^2$ is a proportionality constant. We demonstrate that the momentum distribution can be tailored to completely determine $\rho$ by choosing target POVMs that are informationally complete, namely, the MUB and SIC-POVMs. 

Remarkably, this approach naturally extends to the tomography of arbitrary $m$-qubit states by simply measuring the qubits' momentum correlations, with no further modifications to the experimental setup. The joint probability of finding each qubit in the modes ${\ket{k_{n_1},...,k_{n_m}}}$ is then given by 
\begin{equation}
    p_{n_1,...,n_m} = \text{Tr}\left( E_{n_1,...,n_m} \rho \right),
\end{equation}
where $E_{n_1,...,n_m} = E_{n_1}\otimes ... \otimes E_{n_m}$. Should a set \{$E_n$\} be informationally complete in SU(2), as it is in our implementations, then the set $\{E_{n_1,...,n_m}\}$ is also complete in $\text{SU(2)}^{\otimes m}$ \Cite{Roeland-2022}.

\subsection{Unitary process design}
To find a unitary transformation in the form of Eq. \Eqref{eqn:unitary} satisfying the constraints of Eq. \Eqref{eqn:constraint}, we start by decomposing the position-dependent operator $\mathcal{U}(x)$ in terms of the generators of SU(2):
\begin{equation}
\mathcal{U}(x)=u_0(x)\sigma_0-i(u_1(x)\sigma_1+u_2(x)\sigma_2+u_3(x)\sigma_3),
\label{eqn:pauli}
\end{equation}
where $\lbrace u_i(x)\rbrace$ are real-valued, $\Lambda$-periodic functions, $\sigma_0$ is the $2\times2$ identity operator, and $(\sigma_1,\sigma_2,\sigma_3)$ are the Pauli matrices. Special unitarity implies that
\begin{equation}
u_0(x)^2+u_1(x)^2+u_2(x)^2+u_3(x)^2=1
\label{eqn:uconstraint}
\end{equation}
at each transverse position $x$. Exploiting the spatial periodicity, we expand each function $u_i(x)$ as a Fourier series:
\begin{equation}
u_i(x)=a_{i,0}+\sum_{n=1}^N a_{i,n}\cos({nk_0x})+b_{i,n}\sin({nk_0x}),
\label{eqn:fourier}
\end{equation}
where $N$ is the maximum number of spatial frequencies needed to describe $u_i(x)$. By decomposing the operators $V_n$ analogously to Eq. \Eqref{eqn:pauli},
\begin{equation}
V_n= v_{0,n}\sigma_0 - i (v_{1,n}\sigma_1 + v_{2,n}\sigma_2 + v_{3,n}\sigma_3),
\end{equation}
and using Eq. \Eqref{eqn:kraus}, we obtain
\begin{equation}
    v_{i,n}  = \frac{a_{i,n} + i b_{i,n}}{2},\label{eqn:solution}
\end{equation}
for a subset of modes ${1\leq n \leq n_{\text{max}}}$, for which we impose the set of constraints of Eq. \Eqref{eqn:constraint}. This analytically determines $n_\text{max}$ Fourier coefficients. The coefficients associated with the remaining modes, $n=0$ and ${n_\text{max} < n \leq N}$, are determined by numerically minimizing the following cost function:
\begin{equation}
\mathcal{L}=\sum_{j=1}^X \sum_{i=0}^3 \bigl(u_i({x_j})^2-1\bigr)^2,
\end{equation}
expressing the unitarity condition of Eq. \Eqref{eqn:uconstraint} at each discretized position ${x_j=j\Lambda/X}$. These coefficients correspond to spatial modes that are not associated with a designed $E_n^{\text{(target)}}$, but are nevertheless necessary to satisfy Eq. \Eqref{eqn:uconstraint}. Although their presence results in an overall decrease in the platform efficiency, this effect can be mitigated by exploiting two key symmetries.
First, the realness of $u_i(x)$ implies
$v_{i,-n}=v_{i,n}^{*}$. Since the target POVM elements are projective, the coefficients of the corresponding $V_n$ can be chosen to be real up to a common global phase. Accordingly, $V_n$ and $V_{-n}$ differ only by an
irrelevant global phase, thereby defining the same POVM element:
$E_n=E_{-n}$. This creates a second copy of the set $\{E^{\text{(target)}}_n\}$ corresponding to negative modes, which can also be used for tomography. Second, if the projection onto the orthogonal state $\ket{\Pi_\perp}$ remains accessible, for instance, by using a polarization grating, the elements $\lbrace{E_n^\perp\rbrace}$ are the complementary projectors on the Bloch sphere and can also be used to perform an equivalent tomography in the complementary basis. By leveraging these two symmetries, one can access $4^m$ equivalent POVMs, from which multiple state tomographies can be extracted in parallel. This unique feature can be used to average over experimental imperfections while maximizing the amount of tomographically useful light.
Still, it is preferable to minimize the amount of light coupled to extraneous momentum modes, which itself is a complex multi-parameter optimization problem discussed in previous works \Cite{doi:10.1126/science.aat8196}. In this paper, we adopt a straightforward variational approach, in which we perform consecutive optimizations for fixed numbers of extraneous modes, ${n_\text{ext}=N-n_\text{max}+1}$, and stop the routine once convergence is achieved at an acceptable efficiency threshold, defined as the ratio between the amount of light that can be used for tomography and the total intensity. In the next sections, we provide quantitative details based on the optical implementation and the particular choice of the tomographic states.  
\begin{figure*}[t!]
    \centering
    \includegraphics[width=\linewidth]{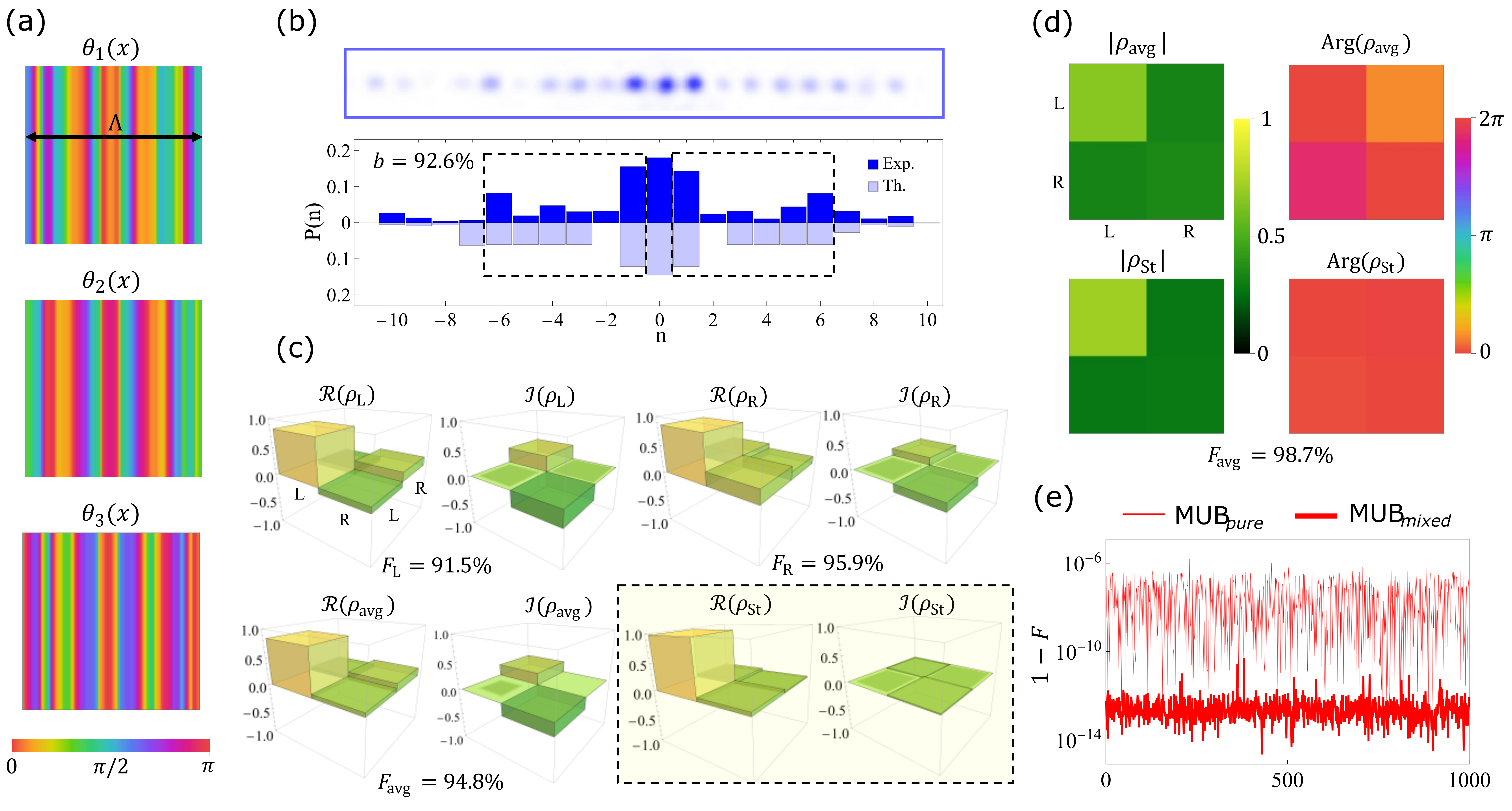}
    \caption{\textbf{MUB tomography with liquid-crystal metasurfaces.} (a) Liquid-crystal patterns for the implementation of single-acquisition tomography with MUB states. (b) From the output diffraction pattern, an experimental probability distribution $P(n)$ is extracted. The probability of designed modes is proportional to the outcome of a set of tomographic measurements. In the example, an input state $\ket{L}$ was considered. (c) The symmetry of the pattern can be used to extract two independent \qo{left} (${\rho_\text{L}}$) and \qo{right} (${\rho_\text{R}}$) tomographies, from which an average density matrix (${\rho_\text{avg}}$) is obtained. All tomographic reconstructions are compared with the result of standard Stokes polarimetry (${\rho_\text{St}}$), reported in the dashed box. (d) Average density matrix obtained for the mixed state ${(\ketbra{L}+\ketbra{H})/2}$. (e) State infidelities obtained from numerical experiments run over 1000 randomly generated pure and mixed states.}
    \label{fig:fig2}
\end{figure*}

\begin{figure*}[ht!]
    \centering
    \includegraphics[width=\linewidth]{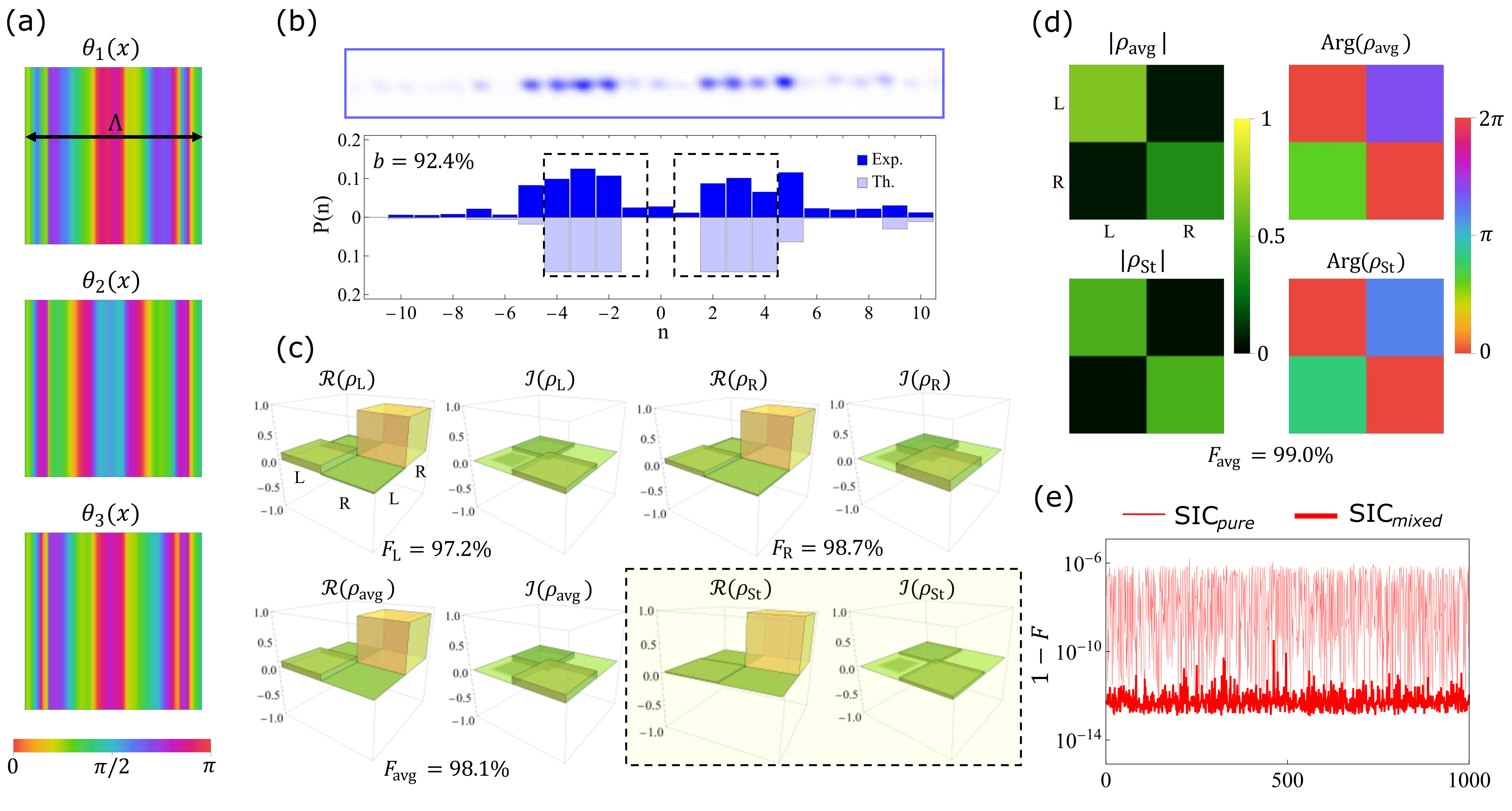}
    \caption{\textbf{SIC-POVM tomography with liquid-crystal metasurfaces.} (a) Liquid-crystal patterns for the implementation of single-acquisition tomography with SIC-POVM states. (b) From the output diffraction pattern, an experimental probability distribution $P(n)$ is extracted. In the example, an input state $\ket{R}$ was considered. (c) The symmetry of the pattern can be used to extract two independent \qo{left} (${\rho_\text{L}}$) and \qo{right} (${\rho_\text{R}}$) tomographies, from which an average density matrix (${\rho_\text{avg}}$) is obtained. All tomographic reconstructions are compared with the result of standard Stokes polarimetry (${\rho_\text{St}}$), reported in the dashed box. (d) Average density matrix obtained for the maximally mixed state ${(\ketbra{L}+\ketbra{R})/2}$. (e) State infidelities obtained from numerical experiments run over 1000 randomly generated pure and mixed states.}
    \label{fig:fig3}
\end{figure*}

\subsection{Optical implementation}
State-of-the-art liquid-crystal technology offers a practical solution to implement complex polarization transformations in the form of Eq. \Eqref{eqn:unitary}. In the basis of circular polarizations, the Jones matrix of a liquid-crystal metasurface can be written as
\begin{equation}
    W(\delta, \theta) = 
    \begin{pmatrix}
    \cos\!\left(\frac{\delta}{2}\right) 
    & i\, e^{-2i\theta(x,y)}\, \sin\!\left(\frac{\delta}{2}\right) \\[6pt]
    i\, e^{2i\theta(x,y)}\, \sin\!\left(\frac{\delta}{2}\right) 
    & \cos\!\left(\frac{\delta}{2}\right)
\end{pmatrix},
\label{eqn:Jones}
\end{equation}
where $\mathcal{\delta} $ is the optical retardation, which is uniform across the metasurface but can be electrically tuned \Cite{Piccirillo2010}, and $\theta (x,y)$ is the spatially varying optic-axis pattern, which is generally two-dimensional. It has been demonstrated that a minimal sequence of three liquid-crystal metasurfaces, $W_1$-$W_2$-$W_3$, tuned so that ${\delta_1=\delta_3=\pi/2}$ and ${\delta_2=\pi}$, can always be found to implement an arbitrary space-dependent polarization transformation \Cite{DiColandrea2023,ammendola2024large}. After individual polarization transformations $\mathcal{U}(x_j)$ have been determined for a target set of projection states (see Eq. \Eqref{eqn:solution}), we map these point-wise operations into the optic-axis patterns of the metasurfaces to be fabricated, ${\lbrace\theta_1(x),\theta_2(x),\theta_3(x)\rbrace}$, employing the analytical approach presented in Refs. \Cite{DiColandrea2023,ammendola2024large}. 

The required optic-axis patterns $\theta_i(x)$, corresponding to the local orientation of the liquid-crystal molecular director in the transverse $xy$ plane, are prepared using a well-established photoalignment technique \Cite{Rubano:19}. The spatial period $\Lambda$ of the complex transformation is set to 2.5~mm. Following the optimization procedure, the photoalignment is also discretized in steps of ${\Delta x=4\,\mu}$m, yielding a total of 625 discrete steps per period. Once the pattern is imprinted, nematic liquid crystals are injected into the sample and naturally align with the dye substrate. The birefringence parameter $\delta$ corresponds instead to the out-of-plane tilt angle of the liquid-crystal molecules, controlled by applying an AC field across the sample. A detailed description of the fabrication process and the characterization of the liquid-crystal electrical response is provided in the Supplementary Material. Figure \figref{fig:fig1}(a) reports a photographic image of two prototypical metasurfaces observed between crossed polarizers, which convert the optic-axis pattern into an intensity distribution. In particular, the image refers to the first metasurface of the sets fabricated for implementing parallel projections onto MUB and SIC-POVM states, respectively.  

\section{Results}
A sketch of the experimental setup for the single-qubit case is shown in Fig.~\figref{fig:fig1}(b). An unknown polarization state, either pure or mixed, is coupled to a Gaussian beam profile with beam waist ${w_0\simeq\Lambda}$, obtained by spatially filtering an 810-nm input laser diode through a single-mode fiber (not shown in the figure). The beam propagates through the three-metasurface gadget, which implements the target space-dependent polarization transformation. Importantly, the three plates must be closely stacked so that free-space propagation between them can be safely neglected. To simultaneously sort the left- and right-handed circular projections, one could employ a polarization grating, such as a liquid-crystal metagrating \Cite{DErrico2020}, that deflects the two orthogonal components symmetrically along $y$. This is illustrated in Fig. \figref{fig:fig1}(c) for the representative case of the six MUB states. In the focal plane of a lens, the momentum modes associated with projections onto these states are symmetrically placed along $x$, while the complementary patterns along $y$ provide access to the orthogonal projections, ultimately enabling the extraction of four independent tomographies from a single experimental acquisition. Here, $\ket{L}$ and $\ket{R}$ are left- and right-handed circular polarization states, ${\ket{H}=(\ket{L}+\ket{R})/\sqrt{2}}$ and ${\ket{V}=(\ket{L}-\ket{R})/\sqrt{2}}$ are horizontal and vertical states, and ${\ket{D}=(\ket{L}+i\ket{R})/\sqrt{2}}$ and ${\ket{A}=(\ket{L}-i\ket{R})/\sqrt{2}}$ are diagonal and anti-diagonal states, respectively. 

In the first experiment, we target the tomographic set of MUB states ${\lbrace{\ket{L},\ket{R},\ket{H},\ket{V},\ket{D},\ket{A} \rbrace}}$, corresponding to modes ${n=1,2,3,4,5,6}$. As mentioned above, under the analytical constraints given by Eq. \Eqref{eqn:solution}, numerical optimization is performed for varying numbers of spatial modes $N$. The outcome was considered satisfactory for ${N=11}$, yielding a theoretical efficiency of ${\eta_\text{MUB}\simeq 77\%}$, corresponding to light used for all four possible tomographies. For simplicity, our experiment used only a single projection onto $\ket{L}$ by cascading a quarter-wave plate and a linear polarizer, thus reducing the actual experimental efficiency to ${\eta_\text{MUB}/2}$. The periodic patterns of the three liquid-crystal metasurfaces resulting from the optimization procedure are shown in Fig. \figref{fig:fig2}(a). As an example, we prepare and characterize an input $\ket{L}$-polarized state. The output diffraction pattern recorded on the camera is shown in the top inset of Fig. \figref{fig:fig2}(b). As expected, the brightest peaks appear at the first order, revealing the projection onto the same state, with very little light coupled to the second order, associated with the orthogonal $\ket{R}$ state. The experimental image is processed to extract a probability distribution (shown in the bottom inset), obtained by integrating the light intensity within each spot and normalizing it to the full spectrum. The agreement between the experimentally reconstructed distribution and the theoretical prediction is quantified through the Bhattacharyya coefficient, $b=\sum_n \sqrt{P_\text{exp}(n)P_\text{th}(n)}$, where the sum extends to all the modes. For this realization, we obtain ${b=92.6\%}$. Two tomographies are extracted by separately processing the negative (${-6\leq n\leq -1}$) and positive (${1\leq n\leq 6}$) modes, enclosed by the black dashed rectangles in Fig. \figref{fig:fig2}(b), hereafter referred to as \qo{left} and \qo{right} tomography. Real and imaginary parts of the reconstructed left and right density matrices, $\rho_\text{L}$ and $\rho_\text{R}$, are reported in Fig. \figref{fig:fig2}(c). By linearly combining the results of the two tomographies, we reconstruct an average density matrix, $\rho_\text{avg}$, also shown in Fig. \figref{fig:fig2}(c). For each reconstruction, we compute the quantum state fidelity, ${F(\rho_1,\rho_2)=\text{Tr}(\sqrt{\sqrt{\rho_1}\rho_2\sqrt{\rho_1}})}$ \Cite{nielsen2010quantum}, with respect to the density matrix obtained from standard Stokes polarimetry of the input state, ${\rho_\text{St}}$, which requires six consecutive projective measurements \Cite{PhysRevA.64.052312}. The fidelities obtained for the left, right, and average density matrices are $91.5\%$, $95.9\%$, and $94.8\%$, respectively, as also reported in Fig. \figref{fig:fig2}(c). 

We also validate the setup using partially polarized input states. These states are prepared synthetically by recording the diffraction patterns for two pure states, processing them as described above, and then summing the two probability distributions with weights set by the target incoherent mixture. Specifically, we target the mixed state ${\rho=(\ketbra{L}+\ketbra{H})/2}$. The tomographic reconstructions obtained for $\ket{H}$ and the resulting mixed state, together with the corresponding probability distribution, are provided in the Supplementary Material. In Fig. \figref{fig:fig2}(d), we compare the experimentally reconstructed density matrix (amplitude and phases), obtained by averaging over the left and right tomographies, with the one obtained via successive Stokes measurements, reporting $F=98.7\%$. 

\begin{figure*}[t!]
    \centering
    \includegraphics[width=0.98\linewidth]{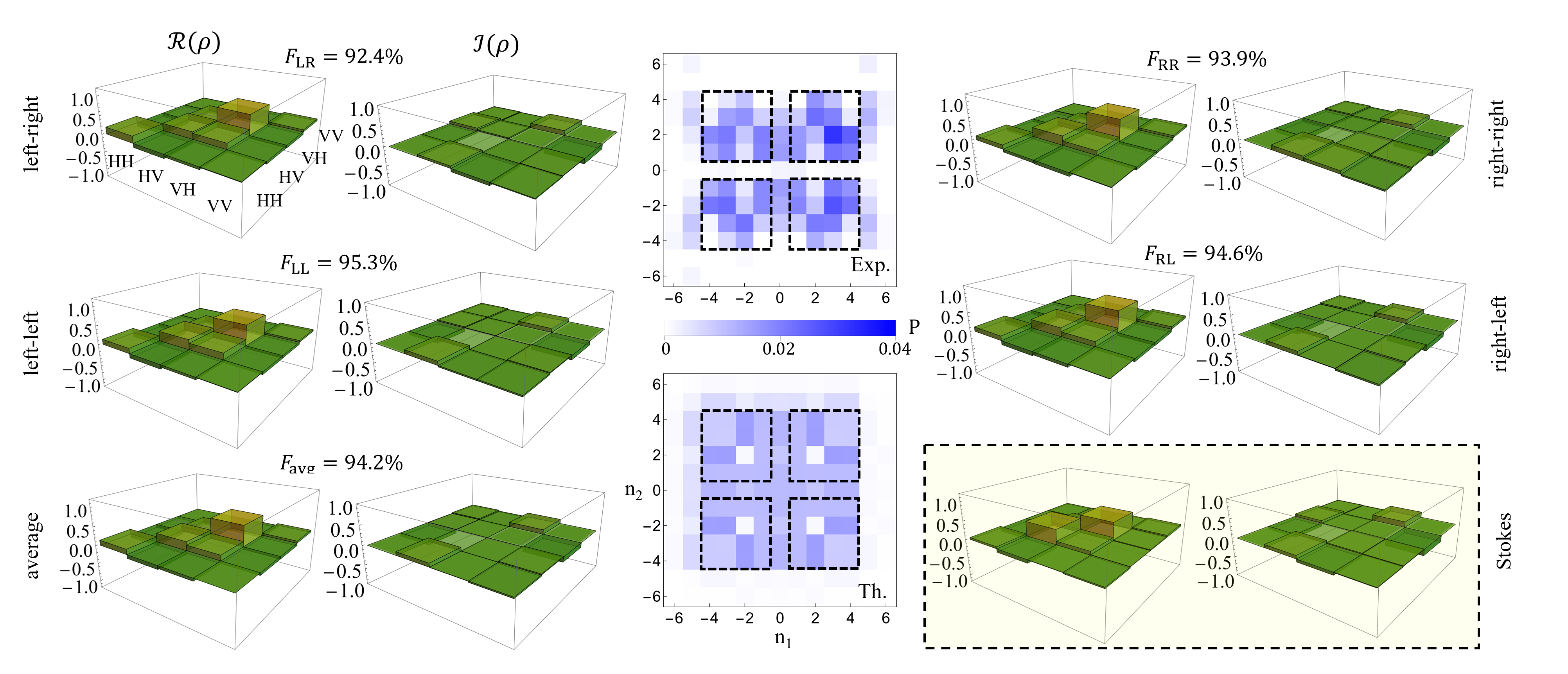}
    \caption{\textbf{Two-photon quantum state tomography.} The set of metasurfaces implementing single-acquisition SIC-POVM tomography is employed with two-photon input states. A time-stamping camera, placed in the far field, enables detection of space-resolved coincidence events. Upon a single polarization projection, four reconstructions of the state are obtained in parallel by recording coincidence events between the \qo{left} and \qo{right} orders of each output diffraction pattern. Experimental reconstructions are compared with standard two-photon Stokes state tomography, reported in the dashed box, prescribing 16 measurements in total.}
    \label{fig:fig4}
\end{figure*}
Figure \figref{fig:fig2}(e) reports the state infidelities, ${1-F}$, retrieved from numerical experiments over 1000 random pure and mixed states. Pure states are generated by uniformly sampling points on the surface of the Bloch sphere. Starting from the set of pure states, ${\lbrace \rho_{\text{pure}}\rbrace}$, mixed states are generated by adding a random depolarization to each state, ${\rho_\text{mixed}=d\,\sigma_0+(1-d)\rho_\text{pure}}$, where $d$ is uniformly sampled in $[0,1]$. Excellent reconstructions are obtained in all cases, with fidelities generally higher for mixed states, which lie closer to the origin of the Bloch sphere. This certifies the high accuracy of our method and its inherent robustness with respect to the input state.

We then fabricate the three liquid-crystal metasurfaces to implement single-acquisition state tomography with SIC-POVM elements $\{E_n = \ketbra{\alpha_i}{\alpha_i}\}$ for $n=1,2,3,4$, defined as follows \Cite{PhysRevX.5.041006}:
\begin{equation}
\begin{split}
\ket{\alpha_1} &= \ket{L},\\
\ket{\alpha_2} &= \sqrt{\frac{1}{3}}\ket{L} + \sqrt{\frac{2}{3}}\ket{R},\\
\ket{\alpha_3} &= \sqrt{\frac{1}{3}}\ket{L} + \sqrt{\frac{2}{3}} e^{2\pi i/3}\ket{R},\\
\ket{\alpha_4} &= \sqrt{\frac{1}{3}}\ket{L} + \sqrt{\frac{2}{3}} e^{4\pi i/3}\ket{R}.
\label{eqn:sicpovms}
\end{split}
\end{equation}
The optic-axis modulations resulting from the optimization are shown in Fig. \figref{fig:fig3}(a), yielding a nominal efficiency of ${\eta_\text{SIC}\simeq 78\%}$. For this experiment, we start with the input state ${\ket{R}}$. The output diffraction pattern is shown in Fig. \figref{fig:fig3}(b). The selected input state is orthogonal to $\ket{\alpha_1}$ (see Eq. \Eqref{eqn:sicpovms}), so no light is expected at the first order, which is only partially realized in the experiment. Deviations from the ideal pattern are mainly ascribed to imperfect metasurface alignment and fabrication defects. For this realization, we obtain ${b=92.4\%}$. Two independent tomographic reconstructions, obtained by separately processing modes ${-4\leq n\leq -1}$ and ${1\leq n\leq 4}$, are reported in Fig. \figref{fig:fig3}(c), from which we also extract an average density matrix. Obtained state fidelities are $97.2\%$, $98.7\%$, and $98.1\%$ for the left, right, and average reconstructions, respectively. 
We also perform quantum state tomography on a maximally mixed state, ${\rho=(\ketbra{L}+\ketbra{R})/2}$. The complete experimental reconstructions are provided in the Supplementary Material. In Fig. \figref{fig:fig3}(d), the average density matrix of the mixed state is compared with the result obtained from Stokes polarimetry, yielding ${F=99\%}$. Moreover, we numerically verify the robustness of the SIC-POVM setup with respect to the input state, reporting the obtained infidelities in Fig. \figref{fig:fig3}(e) for the same set of pure and mixed states considered in Fig. \figref{fig:fig2}(e).

Finally, we validate our device in a genuinely quantum regime, characterizing a two-photon state described by the density matrix
\begin{equation}
\rho = \frac{1}{2}\left(|H_1V_2\rangle\langle H_1V_2| + |V_1H_2\rangle\langle V_1H_2|\right).
\end{equation}
The two-photon mixed state is prepared by first generating polarization-entangled photon pairs through spontaneous parametric down-conversion (SPDC), followed by dephasing induced by random birefringent phase drift in single-mode fibers. The photons exiting the two fibers are sent through the three-metasurface device, but with the two photons vertically displaced, so that the two diffraction patterns along $x$ are imaged at different positions along $y$. Coincidence events between the two diffraction patterns are recorded with a time-stamping intensified camera (TPX3CAM), featuring nanosecond time resolution on each individual pixel \Cite{Nomerotski_2023}. In this way, a single polarization projection provides access to up to four parallel state tomographies. Details of the state preparation are provided in the Supplementary Material. 

The experimentally reconstructed and theoretically predicted two-photon probability distributions obtained from the SIC-POVM three-metasurface platform are shown in Fig.~\figref{fig:fig4}. The measured distribution shows good qualitative agreement with the theoretical expectation, yielding ${b=84\%}$. The observed deviations likely arise from experimental imperfections that cause the prepared state to differ from the ideal mixture. This is supported by comparing the two-photon density matrix reconstructed from our metasurfaces with that obtained using standard Stokes polarimetry, which requires 16 polarization projections for two qubits~\Cite{PhysRevA.64.052312,ALTEPETER2005105}. This comparison yields an average state fidelity of $94.2\%$, obtained from the four separate reconstructions. The obtained purities, computed as $\text{Tr}(\rho^2)$, are ${53.5\%}$ and ${51.6\%}$ for the average and the Stokes reconstruction, respectively. These results confirm that our setup can also be used to efficiently characterize multi-photon states.

\section{Conclusions}
We demonstrated a liquid-crystal platform that enables robust tomography of an arbitrary number of polarization qubits within a single acquisition 
and does not require reconfiguring the measurement apparatus at any stage. The method can be adapted to arbitrary choices of tomographic states. In this work, we demonstrated this flexibility by fabricating metasurfaces that implement polarization state tomography using two representative sets of projective measurements. A key strength of our approach is that the correspondence between an informationally complete POVM and a subset of diffraction orders is derived analytically, and numerical optimization enforces that the overall transformation remains unitary. 
Another unique feature of our scheme is that the same gadget can be used to characterize an arbitrary number of input qubits by exploiting the $y$ direction as a multi-port input channel, with experimental limitations set by the overall setup efficiency. The latter also depends on the transmittance of individual metasurfaces, measured to be around $80\%$, but this could be significantly improved with suitable anti-reflection coatings. Finally, we leveraged the natural symmetry of the diffraction patterns to extract multiple experimental tomographic reconstructions of the input state from a single acquisition. This becomes especially relevant in the multi-photon regime, where redundant tomographic information helps mitigate the exponential increase of losses. 

Several extensions of this work could be pursued in the future. This setup could be adapted to implement single-shot partial tomography from informationally incomplete measurements, where knowledge of the state is considered satisfactory above a given threshold \Cite{binosi2024tailormade,Zambrano2024certificationof}. In analogy to shadow tomography \Cite{doi:10.1137/18M120275X}, the optimization of the metasurfaces could be carried out by targeting only specific properties of the system, such as entanglement witnesses \Cite{ziaScience}, without fully reconstructing the unknown state. 
While in this work we retrieved the qubit state by coupling it to the light transverse momentum, it would be interesting to generalize the metasurface approach for the tomography of a second high-dimensional spatial degree of freedom, such as the orbital angular momentum \Cite{wang2023characterization,niu2025randomly,Shafran:26}. For this task, the increased computational complexity could be tackled with neural networks \Cite{PhysRevLett.126.170504,chen2023neural}, trained directly on raw experimental data \Cite{Pierangeli2023,PhysRevLett.132.160802}. For the tomography of high-dimensional spatial qudits, a recent study has identified randomly structured metasurfaces as a promising resource-effective alternative to traditional inverse-design approaches, with intrinsic tolerance to fabrication errors \Cite{niu2025randomly}. This feature could also be explored using liquid crystals, which can serve as a random-scattering medium \Cite{LCbranching1,lcbranching2}. Finally, we plan to extend our study to quantum process tomography of polarization optics \Cite{DiColandrea:23_qpt}, which is experimentally mapped onto single-shot imaging of Mueller matrices \Cite{zaidi2024metasurface}.

\section*{Acknowledgements}
This work was supported by the Canada Research Chairs (CRC), the NRC Quantum Sensing Programme, Quantum Enhanced Sensing, Imaging (QuEnSI) Alliance Consortia Quantum grant. FDC acknowledges financial support from the European Union–Next Generation EU through the PNRR MUR Project No.~PE0000023-NQSTI. 

\noindent 
\newpage
\bibliography{main}

\clearpage
\onecolumngrid
\renewcommand{\figurename}{\textbf{Figure}}
\setcounter{figure}{0} \renewcommand{\thefigure}{\textbf{S{\arabic{figure}}}}
\setcounter{table}{0} \renewcommand{\thetable}{S\arabic{table}}
\setcounter{section}{0} \renewcommand{\thesection}{S\arabic{section}}
\setcounter{equation}{0} \renewcommand{\theequation}{S\arabic{equation}}
\onecolumngrid

\begin{center}
{\Large Supplementary Material for: \\Single-acquisition tomography of photonic qubits with structured media}
\end{center}
\vspace{1 EM}
\section{Fabrication and characterization of liquid-crystal metasurfaces}
\begin{figure*}[h!]
    \centering
    \includegraphics[width=0.65\linewidth]{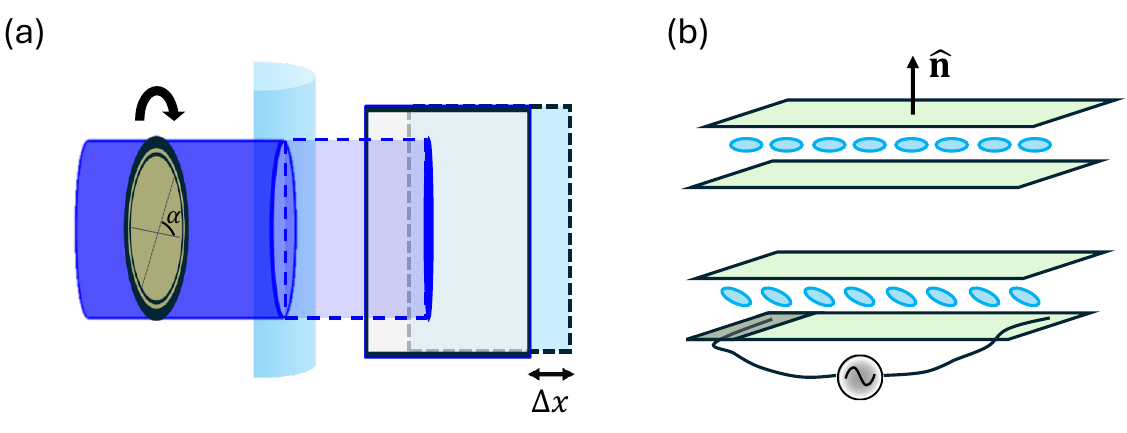}
    \caption{\textbf{Photoalignment and electrical tunability of liquid-crystal metasurfaces.} (a) The sample is translated along the patterning direction in steps of ${\Delta x=4\,\mu}$m. A cylindrical lens focuses the polarized beam at 445 nm along $x$, since the required modulations are one-dimensional by design. At each position, the azo-dye substrate aligns with the incident linear polarization, controlled by a rotating half-wave plate. At the end of the procedure, nematic liquid crystals are injected and penetrate the sample by capillarity. (b)~The application of an external field torques the liquid-crystal molecules towards the cell normal vector $\hat{\textbf{n}}$, thus modifying the effective birefringence of the device.}
    \label{fig:fabrication}
\end{figure*}

Liquid-crystal metasurfaces are birefringent slabs of nematic liquid crystals, whose molecular director is patterned on the micrometric scale. The fabrication process starts by cleaning two glass plates, precoated with a thin film of indium tin oxide (ITO), a transparent conductive material, in an ultrasonication chamber, first with a 5\% solution of phosphate-free alkaline detergent and distilled water, then with distilled water only. Both cleaning steps are performed at ${60^\circ \text{C}}$ for 1 hour. The plates are then dried in an oven at $100^\circ\text{C}$ for 1 hour. The glass plates are spin-coated with a 0.1\% solution of Brillant Yellow, a photosensitive azo dye, dissolved in Dimethylformamide (DMF). After spin-coating, the plates are sandwiched together using $4$-$\mu$m silica spacers, which sets the cell thickness. The desired pattern is imprinted using a well-established photoalignment technique~\cite{Rubano:19}, illustrated in Fig. \figref{fig:fabrication}(a), through which the azo-dye molecules orient point by point following the polarization of an incident linearly polarized beam at $445$~nm. After photoalignment, the cell is filled with a mixture of nematic liquid crystals (6CHBT), which align with the azo-dye substrate and penetrate by capillarity the sample pre-heated at $100^\circ\text{C}$.

The application of an electric field induces an out-of-plane tilt of the liquid-crystal molecules, which is exploited to control the birefringence of the device (see Fig. \figref{fig:fabrication}(b)). In the following, we illustrate the protocol to characterize the electrical response of a liquid-crystal metasurface through polarimetric measurements. In the circular polarization basis, where ${\ket{L}=(1,0)^T}$ and ${\ket{R}=(0,1)^T}$ represent the left-handed and right-handed circular polarization states, respectively, the Jones matrix of a liquid-crystal metasurface reads
\begin{equation}
W(\delta,\theta)=
\begin{pmatrix}
\cos\frac{\delta}{2} & i\sin\frac{\delta}{2}e^{-2i\theta}\\[4pt]
i\sin\frac{\delta}{2}e^{2i\theta} & \cos\frac{\delta}{2}
\end{pmatrix},
\label{eqn:unitarywaveplate}
\end{equation}
where $\delta$ is the birefringence parameter, which is uniform across the sample but can be electrically controlled, and $\theta$ is the optic-axis orientation. At a given voltage $V$, the birefringence parameter $\delta$ is extracted by illuminating the sample with a $\ket{L}$-polarized input beam and projecting the transmitted light on the $\ket{L}$ and $\ket{R}$ polarizations. The value of $\delta$ is obtained as
\begin{equation}
\delta(V)=2\arctan\sqrt\frac{I_{LR}(V)}{I_{LL}(V)},
\label{eqn:delta}
\end{equation}
where ${I_{LR}}$ (${I_{LL}}$) is the intensity of the right (left) projection, corresponding to the converted (unconverted) fraction of incoming light \Cite{Piccirillo2010}.

\section{Two-photon experimental setup}

\begin{figure*}[t!]
    \centering
    \includegraphics[width=0.9\linewidth]{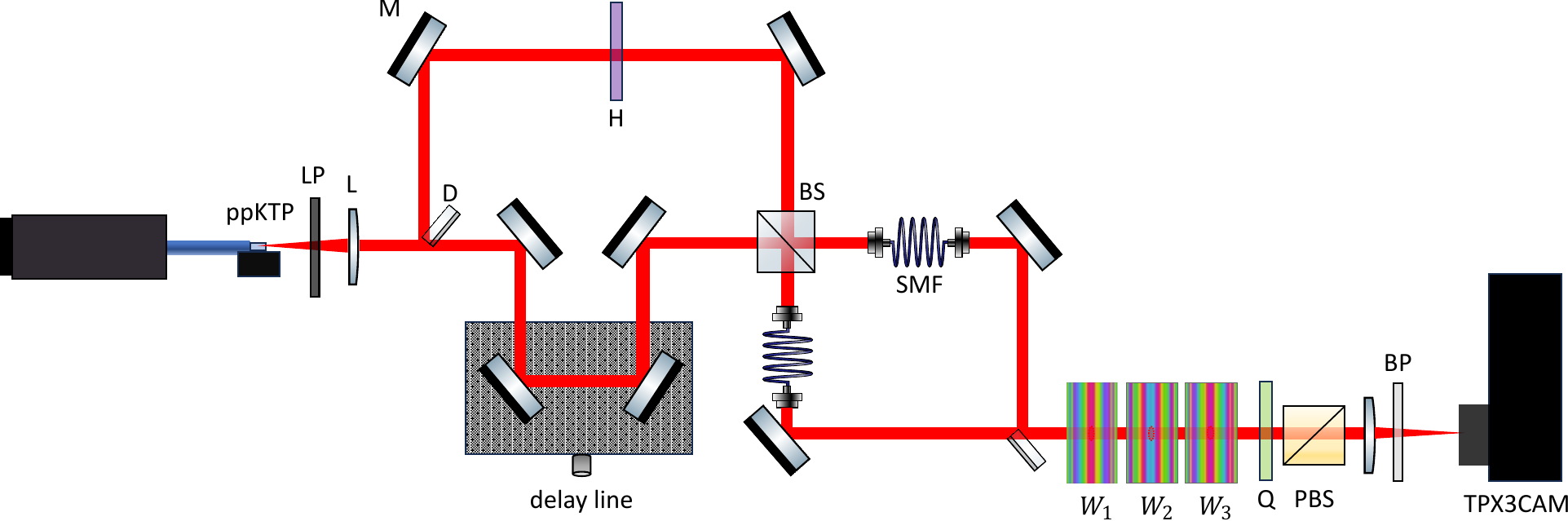}
    \caption{\textbf{Two-photon experimental setup.} A 405-nm continuous-wave laser pumps a Type-0, 1-mm-long ppKTP crystal, generating photon pairs sharing the same polarization state. The polarization of one of the two is rotated through a half-wave plate (H). The photons enter a BS and, at the output, they are coupled to SMFs for spatial-mode filtering. A delay line is used to control the temporal overlap of the two photons. The photons pass through the three metasurfaces (${W_1}$-${W_3}$), while a small vertical tilt is applied to one input, followed by a global polarization projection (Q-PBS). At the output, a lens implements an all-optical Fourier transform, converting momentum modes into transverse positions in the focal plane, where an intensified single-photon camera (TPX3CAM) enables detection of space-resolved coincidence events. L: Lens, LP: Longpass filter, BP: Bandpass filter, BS: Beamsplitter, PBS: Polarizing beamsplitter, SMF: Single-mode fiber, H: Half-wave plate, Q: Quarter-wave plate, M: Mirror, D: D-shaped mirror.}
    \label{fig:figS1}
\end{figure*}

The experimental setup used to generate and characterize two-photon states is plotted in Fig. \figref{fig:figS1}. Photon pairs are generated through spontaneous parametric down-conversion (SPDC) in a type-0, 1-mm-long periodically poled potassium titanyl phosphate (ppKTP) crystal in the state $|\Psi\rangle_{1,2}=|H_1H_2\rangle$, where the subscripts $1,2$ denote the paths of the two photons. The polarization of one photon is rotated from $\ket{H}$ to $\ket{V}$ via a half-wave plate, and the two photons are then sent onto a 50:50 beamsplitter (BS) to undergo Hong-Ou-Mandel interference. Conditioned on detecting one photon in each output port of the BS, the measured two-photon state is
\begin{equation}
|\Psi^{-}\rangle_{1',2'}= \frac{1}{\sqrt{2}}\left(|H_{1'}V_{2'}\rangle-|V_{1'}H_{2'}\rangle\right),
\end{equation}
where $1',2'$ denote the two output ports of the BS. The photons from the two output ports are then coupled to separate single-mode fibers (SMFs). Due to fiber birefringence, the horizontal and vertical polarization components acquire different phases,
\begin{equation}
|H_i\rangle \rightarrow e^{i\phi_{H_i}}|H_i\rangle,
\qquad
|V_i\rangle \rightarrow e^{i\phi_{V_i}}|V_i\rangle.
\end{equation}
For a fixed birefringent phase, the state after the fibers can be written as
\begin{equation}
|\Psi(\Delta\phi)\rangle_{1',2'} = \frac{1}{\sqrt{2}} \left(|H_{1'}V_{2'}\rangle - e^{i\Delta\phi}|V_{1'}H_{2'}\rangle\right),
\end{equation}
where
\begin{equation}
\Delta\phi = (\phi_{V_1}+\phi_{H_2}) - (\phi_{H_1}+\phi_{V_2}) .
\end{equation}
In practice, the birefringent phase drifts during the measurement due to temperature fluctuations and mechanical stress in the fibers. Since the experimentally reconstructed state is averaged over the acquisition time, this phase drift leads to partial or complete dephasing between the $|H_{1'}V_{2'}\rangle$ and $|V_{1'}H_{2'}\rangle$ components. The resulting two-photon state is therefore described by the mixed density matrix
\begin{equation}
\rho_{1',2'} = \frac{1}{2} \big(|H_{1'}V_{2'}\rangle\langle H_{1'}V_{2'}| + |V_{1'}H_{2'}\rangle\langle V_{1'}H_{2'}| - \gamma |H_{1'}V_{2'}\rangle\langle V_{1'}H_{2'}| - \gamma^{*} |V_{1'}H_{2'}\rangle\langle H_{1'}V_{2'}|\big),
\end{equation}
where $\gamma=\left\langle e^{-i\Delta\phi}\right\rangle$ quantifies the residual coherence after averaging over the phase drift. In the limit of a stable birefringent phase, $|\gamma|=1$, the state remains a pure maximally entangled state up to a relative phase. In the opposite limit, where the phase is fully randomized during the acquisition, $\gamma=0$, and the state reduces to the mixed state
\begin{equation}
\rho_{1',2'} = \frac{1}{2} \big(|H_{1'}V_{2'}\rangle\langle H_{1'}V_{2'}| + |V_{1'}H_{2'}\rangle\langle V_{1'}H_{2'}|\big).
\end{equation}

The photons exiting the two fibers are sent through the three-metasurface device, and the resulting diffraction pattern is imaged using a time-stamping single-photon camera (TPX3CAM) with nanosecond timing accuracy. The two photons are vertically displaced so that they are collected by different regions of the camera. Using the photon arrival-time information recorded by the camera, photon pairs are identified through coincidence measurements. The two-photon spatial correlation between the two diffraction bands is then obtained by plotting the detected spatial coordinates of the paired photons. 

\newpage
\section{Supplementary data}
\begin{figure*}[ht]
    \centering
    \includegraphics[width=\linewidth]{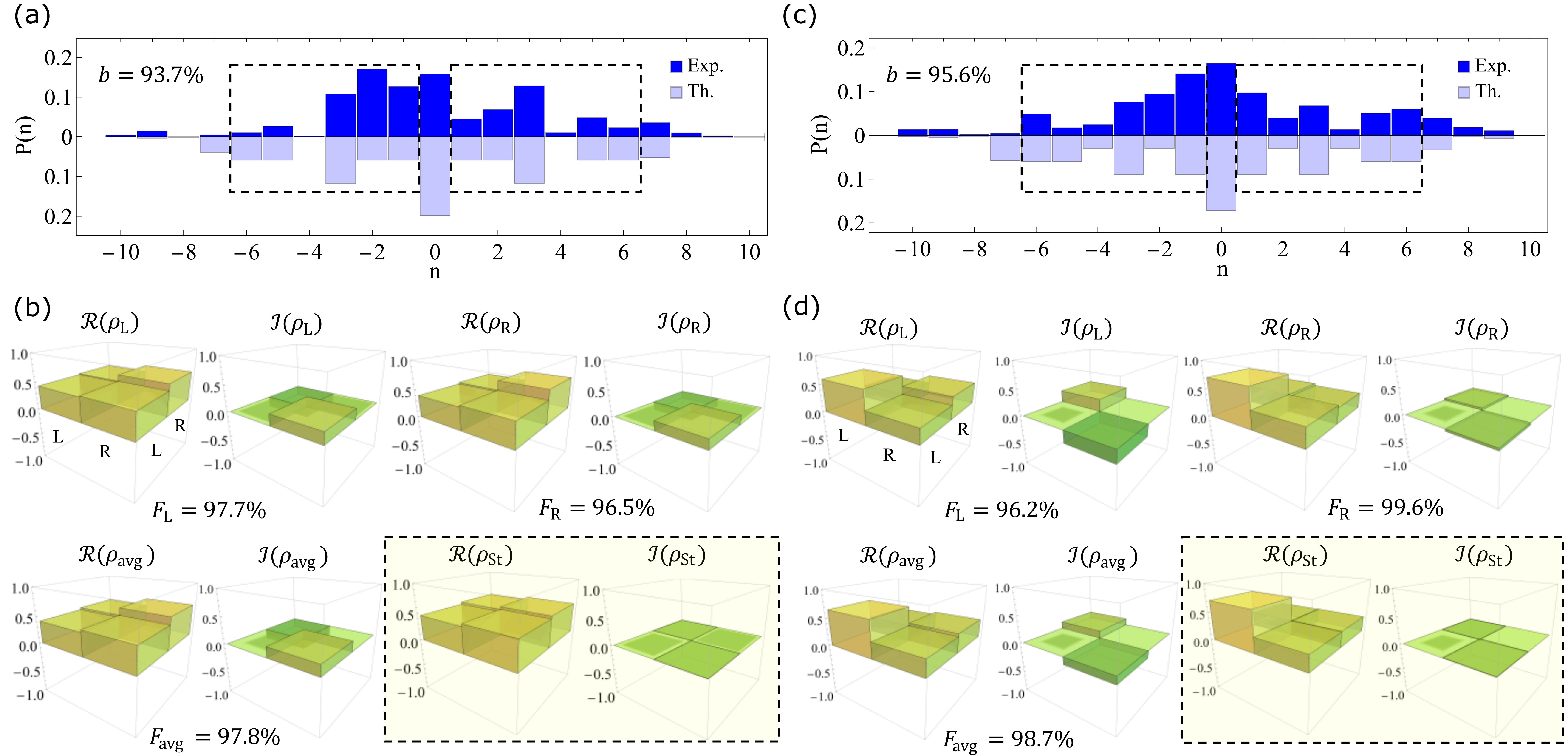}
    \caption{\textbf{MUB tomography with liquid-crystal metasurfaces.} Experimentally reconstructed probability distributions for an input pure (a) $\ket{H}$ and mixed (c)~$(\ketbra{L}+\ketbra{H})/2$ state, obtained from the metasurfaces implementing the MUB tomography, compared with theoretical predictions. The extracted left, right, and average density matrices are shown in (b) and (d), respectively, and compared with Stokes polarimetry, reported in the dashed box.}
    \label{fig:figS2}
\end{figure*}

Figure \figref{fig:figS2} shows the experimentally reconstructed probability distributions for the input state (a) $\ket{H}$ and (c)~$(\ketbra{L}+\ketbra{H})/2$, obtained from single-acquisition MUB tomography, compared with theoretical predictions. The reconstructed left, right, and average density matrices are shown in Fig. \figref{fig:figS2}(b) and (d) for the two cases, respectively, and compared with standard Stokes polarimetry.

\begin{figure*}[ht]
\vspace{0 cm}
    \centering
    \includegraphics[width=\linewidth]{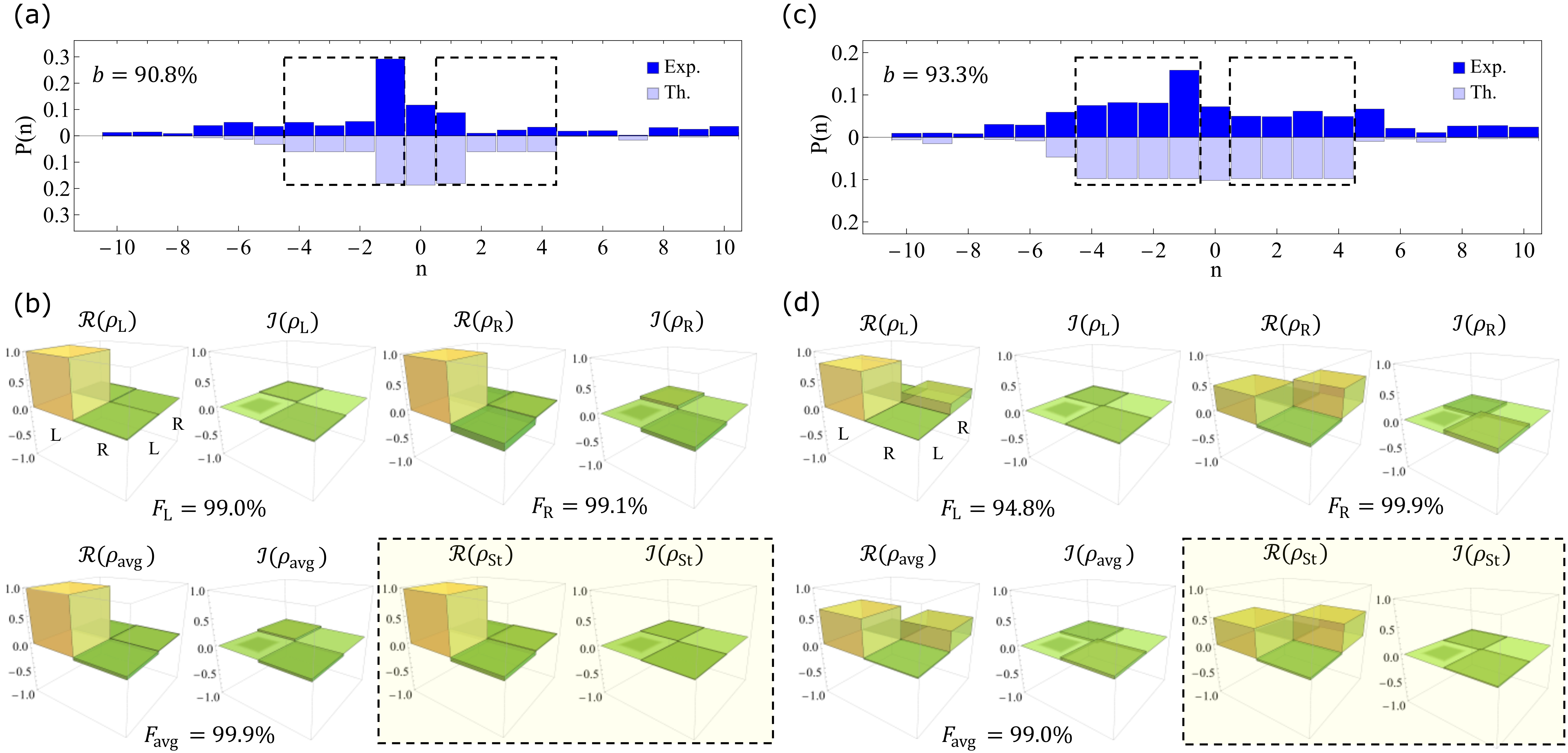}
    \caption{\textbf{SIC-POVM tomography with liquid-crystal metasurfaces.} Experimentally reconstructed probability distributions for an input pure (a) $\ket{L}$ and mixed  (c)~$(\ketbra{L}+\ketbra{R})/2$ state, obtained from the metasurfaces implementing the SIC-POVM tomography, compared with theoretical predictions. The extracted left, right, and average density matrices are shown in (b) and (d), respectively, and compared with Stokes polarimetry, reported in the dashed box.}
    \label{fig:figS3}
\end{figure*}

Figure \figref{fig:figS3} shows the experimentally reconstructed probability distributions for the input state (a) $\ket{L}$ and (c)~$(\ketbra{L}+\ketbra{R})/2$, obtained from single-acquisition SIC-POVM tomography, compared with theoretical predictions. The reconstructed left, right, and average density matrices are shown in Fig. \figref{fig:figS3}(b) and (d) for the two cases, respectively, and compared with standard Stokes polarimetry.

\end{document}